\documentstyle[11pt]{article}
\include{epsf}
\begin{document}
%%%%%%%%%%%%%%%%%%%%%%%%%%%%%%%%%%%%%%%%%%%%%%%%%%%%%%%%%%%%%%%%%%%%%%%%%%%

\thispagestyle{empty}

\begin{flushright}
MIT-CTP-2839 \\ hep-lat/9903021 
\end{flushright}
\begin{center}
\vspace*{5mm} {\LARGE The lattice Schwinger model as a \vskip2mm
discrete sum of filled Wilson loops} 
\vskip10mm \centerline{ {\bf Christof Gattringer}}
\vskip 2mm Massachusetts Institute of Technology \\ Center for
Theoretical Physics \\ 77 Massachusetts Avenue \\ Cambridge MA 02139,
USA \vskip22mm
\begin{abstract}
Using techniques from hopping expansion we identically map 
the lattice Schwinger model with Wilson fermions to a model of
oriented loops on the lattice. This is done by first computing the 
explicit form of the fermion determinant in the external field. 
Subsequent integration of the gauge fields 
renders a sum over all loop configurations with simple Gaussian weights 
depending on the number of plaquettes enclosed by the loops.
In our new representation vacuum expectation values of local fermionic 
operators (scalars, vectors) can be computed by simply counting the  
loop flow through the sites (links) supporting the scalars (vectors).
The strong coupling limit, possible applications of our methods to 4-D models 
and the introduction of a chemical potential are discussed.
\end{abstract}
\end{center}
\vskip6mm
\noindent
{\sl To appear in Nuclear Physics B}
\newpage
\setcounter{page}{1}
\section{Introduction} 
\noindent
The evaluation of the fermion determinant in lattice field theories 
poses a highly non-trivial problem. In particular for an odd number 
of flavors, or when a chemical potential is added, no fast 
numerical algorithms are known.
As a possible way out several studies of alternative representations of the
fermion determinant in terms of fermion loops can be found in the literature
\cite{RoWo84}-\cite{Ga98b}. Loop representations  
are not only useful as a tool for numerical simulations, but
also provide interesting insights into the nature of the fermion determinant,
in particular they provide a solution of the fermion sign 
problem (compare \cite{Cr98,ChWi99} for recent developments).
At least for abelian gauge fields it is furthermore possible to perform 
a duality transformation for the link variables and in 2 dimensions they can 
even be integrated in closed form. 
The resulting representation in 2-D is
a discrete sum over filled Wilson loops.

The historic approach to finding a loop representation for the fermion 
determinant is to first expand the Boltzmann factor for the fermions. Since
the fermions are represented by Grassmann variables the power series for the
Boltzmann factor has only finitely many terms. In a second step the Grassmann
integral at each site is saturated with all possible combinations of the 
terms from the series for the Boltzmann factor. The result can be viewed as 
a sum over closed loops of links. We remark, that for gauge theories products 
of link variables arranged along closed loops are the only possible gauge 
invariant terms and thus structures of closed loops can be expected already 
from gauge invariance.

The local saturation of the Grassmann integral is however a tedious algebraic
problem. Thus most of the results were obtained for staggered fermions where
the additional structure due to spinorial degrees of freedom is avoided 
\cite{RoWo84}-\cite{ArFoGa96}. The resulting model however  
still describes more than one 
fermion due to the doubling problem, which is only
reduced but not completely eliminated by the staggered formulation.

The first result for Wilson fermions is Salmhofer's mapping of the
strongly coupled Schwinger model to a 7 vertex model \cite{Sa91}.
The techniques used there were later generalized by Scharnhorst and applied
to the strongly coupled Schwinger model with 2 flavors and the 2-D lattice
Thirring model \cite{Sch96,Sch97}. Thus for purely fermionic models in 
2-D also loop representations for Wilson fermions have been found for some
models. However, no loop representation for the fermion determinant in an
external field has been found for Wilson fermions while for the staggered 
case such results are known \cite{ArFoGa96}.

In a recent publication we have attacked this problem using a new approach
\cite{Ga98a,Ga98b}. The central idea is to first perform a hopping 
expansion of
the fermion determinant (see e.g.~\cite{MoMu94} for an introduction and
\cite{Lu90} for an application to the Schwinger model). The 
hopping expansion essentially gives a loop representation for the free 
energy of fermions in an external field. Thus it already provides
the natural variables for the loop representation. The remaining problem is
to expand the Boltzmann factor in order to bring down the loops from the
exponent. In \cite{Ga98a,Ga98b} this method has been applied to the case of 
2-D Wilson fermions interacting with an external 
scalar field through a Yukawa coupling.
\\

In this article we treat the case of the lattice Schwinger model with 
Wilson fermions using the hopping expansion technique. The fermion determinant 
in the external U(1) gauge field will be written as a sum over oriented 
loops dressed with the gauge variables sitting on the links of the loops. 
In this representation we then integrate explicitly the gauge fields and the 
partition function turns into a discrete sum over filled Wilson loops. 
Vacuum expectation values of Wilson loop type observables
can easily be taken into account by adding the Wilson 
loops to the loops from the
determinant before integrating the gauge fields. VEV's of local fermionic
observables (scalars and vectors) also turn out to have rather 
simple expressions
in the loop formalism. They can be obtained by differentiating the
fermion determinant (before performing the gauge field integration) 
with respect to the mass term for the scalars or with respect to 
the link variables for the vectors. In the loop picture these observables then 
correspond to the loop flow through sites (for the scalars) 
and links (vectors). Thus
in a numerical simulation of the loop model 
the evaluation of n-point functions
for these observables becomes exceedingly simple. Finally we remark, that a 
chemical potential can be included in a natural way by giving the pieces
of loops running forward in time a weight different from the weight 
for pieces running
backward in time.  

The article is organized as follows: In the next section we rederive the
hopping expansion of the fermion determinant in a form which is convenient
for our later manipulations. In Section 3 we expand the Boltzmann factor and 
construct the loop representation of the determinant. This is done by first
analyzing several examples for different contributions and subsequently the
general rules of the loop calculus are discussed. In the end of Section 3
we also briefly address the role of a chemical potential.
In Section 4 we discuss the
relation of our new representation to known results which can be obtained
as limiting cases of our formulas. In particular the coupling to an 
external Yukawa field and the strongly coupled Schwinger model are considered. 
In Section 5 we integrate out the gauge fields and give the final 
representation for the partition function of the lattice Schwinger model 
with 
Wilson fermions in terms of filled loops.  In Section 6 we discuss the
loop representation of some observables 
and the article ends with a discussion 
in Section 7.

\section{Setting and hopping expansion } 
\noindent
We study the lattice Schwinger model with Wilson fermions. In addition
to the gauge fields we also couple a real scalar field $\theta$ to the
fermions through a Yukawa-type interaction. The scalar field $\theta$
will serve  as a tracking device for the algebraic manipulations below
and also allows for a consistency check by comparing our new
representation with the known loop representation for the Yukawa-type
coupling \cite{Ga98b}. In addition we will use $\theta$ as a source term 
for generating n-point functions. 

The Wilson lattice action is a bilinear form $S = \sum_{x,y}
\overline{\psi}(x) K(x,y) \psi(y)$ with kernel
\begin{equation}
K(x,y) \; = \; \Big[2 + m + \theta(x)\Big] \; \delta_{x,y}  \; - \;
\sum_{\mu = \pm 1}^{\pm 2} \Gamma_\mu U_\mu(x) \;
\delta_{x+\hat{\mu},y} \; ,
\label{kernel}
\end{equation}
where we defined
\[
\Gamma_{\pm \mu} \; = \; \frac{1}{2} [ 1 \mp \sigma_\mu ] \; \; \; \;
, \; \; \; \; \mu = 1,2 \; .
\]
Here $\sigma_1,\sigma_2$ are Pauli matrices. $U_\mu(x) \in$ U(1) are
the link variables  supported on the links $(x,\mu)$ of the lattice,
with $U_{-\mu}(x) = U_{\mu}(x-\hat{\mu})^*, (\mu =1,2)$, i.e.~for
hopping in backward direction the link variables are complex
conjugate. The sum in the bilinear form  runs over the whole lattice,
$x,y \in \Lambda$.  For simplicity we assume that our lattice $\Lambda$ is a
finite rectangular piece of Z\hspace{-1.3mm}Z$^2$ with open boundary
conditions, i.e.~hopping terms that would lead to the  outside of our
lattice are omitted. We remark that all our results also hold for
$\Lambda$ being a cylinder with a compact time direction with
anti-periodic boundary conditions and open boundary
conditions for the space direction. Such a geometry is interesting for 
analyzing the model at finite temperature. Let's define
\[
M(x) \; = \; 2 + m + \theta(x) \; ,
\]
and assume that $\theta(x)$ is such that $M(x) \neq 0$ for all lattice
points $x$. This is a purely technical assumption due to the
particular techniques  we use for computing the determinant. The final
result will be a finite  polynomial in the $\theta(x)$ and the above
restriction is irrelevant then.  We now can write for the fermion
determinant
\begin{eqnarray} 
& & \int d \overline{\psi} d \psi 
\exp\left( - \sum_{x,y \in \Lambda} 
\overline{\psi}(x) K(x,y) \psi(y) \right) \; = \; 
\det K  \nonumber \\
& = & \prod_{x \in \Lambda} M(x)^2 \;  \det\Big(1 - H \Big) 
\; = \; \prod_{x \in \Lambda} M(x)^2 \exp \left( - \sum_{n=1}^\infty
\frac{1}{n} \mbox{Tr} \;  H^n \right) \; .
\label{hopexp}
\end{eqnarray}
In the last step the hopping expansion was performed, i.e. the
determinant  was expressed using the well known trace-logarithm
formula and the logarithm was expanded in a power series.  The hopping
matrix $H$ is defined as
\begin{equation}
H(x,y) \; = \; \sum_{\mu = \pm 1}^{\pm 2} \; \Gamma_{\mu} \;
\frac{U_\mu(x)}{M(x)} \; \delta_{x+\hat{\mu}, y} \; .
\label{hoppingmatrix}
\end{equation}
The series in the exponent of (\ref{hopexp}) converges for  $||H|| <
1$, and the whole expression represents the determinant then 
(see e.g.~\cite{ReSi78}).
In \cite{WeCh79} it is shown for the case of $\theta(x) = 0$, that 
$||H|| < 1$ for $m > 0$. It is straightforward to generalize this argument 
and prove that for our case we have $||H|| < 1$ for $m + \theta(x) > 0$.
To summarize: the series in (2) converges for $m + \theta(x) > 0$ and the 
right hand side of (\ref{hopexp}) represents the determinant for this
range of parameters.
 
Due to the Kronecker delta
in (\ref{hoppingmatrix}), the contributions to Tr$H^n$ are
supported  on closed loops on the lattice, and since closed loops are
of even length (for the cylinder geometry discussed above we assume an even 
length for the compact direction), the contributions for odd $n$ vanish. 
For even $n =
2k$ we obtain
\begin{equation}
\mbox{Tr} \; H^{2k} \; = \;  \sum_{x \in \Lambda} \; \sum_{l \in {\cal
L}^{(2k)}_x} \; \prod_{(y,\mu) \in l} \frac{U_\mu(y)}{M(y)} \; \;
\mbox{Tr} \; \prod_{\nu \in l} \Gamma_\nu \; .
\label{hoptrace}
\end{equation}
Here ${\cal L}^{(2k)}_x$ is the set of all closed, connected loops of
length  $2k$ and base point $x$. In the first product we collect all
the link variables  $U_\mu(y)/M(y)$ on the links visited by the loop
according to their orientation.  The last term in (\ref{hoptrace}) is
the trace of the ordered product of the hopping generators
$\Gamma_\nu$ as they  appear along the loop $l$. We remark, that
$\Gamma_{\pm \nu} \Gamma_{\mp \nu} = 0$, which implies that whenever a
loop turns around at a site and runs back along its last link this
contribution vanishes. Thus all these {\it back-tracking} loops can be
excluded from ${\cal L}_x^{(2k)}$.

Evaluating the trace over the matrices $\Gamma_\nu$  for a given loop
is the  remaining problem in the hopping expansion.  It first has been
solved for the infinite lattice in \cite{St81} by realizing that the
Pauli matrices  give rise to a representation of discrete  rotations
on the lattice. Alternatively one can build up the loops using four
basic steps and compute the trace in  an inductive procedure along the
lines of \cite{GaJaSe98}. The latter method can be extended to the
case of lattices with compact dimensions. 
When choosing a cylinder with anti-periodic boundary conditions
in time direction and open b.c.~for the space direction the result
equals the result for the rectangular lattice with open boundary conditions.
Thus the result for both of the discussed sets of boundary conditions is
\begin{equation}
\mbox{Tr} \; \prod_{\nu \in l} \Gamma_{\nu} \; = \;  -(-1)^{s(l)}
\Big( \frac{1}{\sqrt{2}} \Big)^{c(l)} \; .
\label{gtrace}
\end{equation}
By $s(l)$ we denote the number of self-intersections of the loop $l$
and  $c(l)$ gives its number of corners. 
We remark, that for e.g.~a torus additional sign factors for 
topologically non-trivial loops would appear. 
The result (\ref{gtrace}) is
independent of the orientation of the loop.  Inserting
(\ref{hoptrace}) and (\ref{gtrace}) in (\ref{hopexp}) we obtain
\begin{eqnarray}
\det K\!& = &\! \prod_{x \in \Lambda} M(x)^2 \; \exp \left(
\sum_{k=1}^\infty \frac{1}{2k}  \sum_{y \in \Lambda} \; \sum_{l \in
{\cal L}^{(2k)}_y}  (-1)^{s(l)} \Big(\frac{1}{\sqrt{2}}\Big)^{c(l)} \!
\prod_{(z,\mu) \in l} \frac{U_\mu(z)}{M(z)} \right)  \nonumber \\ & =
&\! \prod_{x \in \Lambda} M(x)^2 \; \exp \left( \sum_{l \in {\cal L}}
\frac{(-1)^{s(l)}}{I(l)} \Big(\frac{1}{\sqrt{2}}\Big)^{c(l)} \!
\prod_{(y,\mu) \in l} \frac{U_\mu(y)}{M(y)} \right) \; .
\label{finalhop}
\end{eqnarray}
In the second step we removed the explicit summation over the base
points.  A loop of length $2k$ without complete iteration of  its
contour allows for $2k$ different choices of a base point thus
canceling the factor $1/2k$ in the previous expression.  A loop which
iterates its whole contour $I(l) > 1$ times allows only for $2k/I(l)$
different base points and a factor $1/I(l)$ remains (compare
\cite{St81,GaJaSe98} for a more detailed discussion). We defined ${\cal L}$ to
be the set of all closed, connected, non back-tracking  loops of
arbitrary length. Note that each loop comes with both of its
orientations and the factors from the link variables
for the two orientations are complex conjugate to each other.

\section{Expansion of the Boltzmann factor and identification of the loop 
representation}
\noindent
The final expression (\ref{finalhop}) for the hopping expansion
performed  in the last section is the starting point for the
identification of the loop representation for the determinant in this
section. As already outlined in the discussion above we will proceed by
expanding the Boltzmann factor. This expansion converges and represents 
the determinant when the sum in the exponent converges, i.e.~for 
$m + \theta(x) > 0$ as discussed above. The resulting series converges 
even absolutely and the reordering of terms which we perform below is 
justified (see e.g.~\cite{He84}).

\begin{figure}[ht]
\centerline{ \epsfysize=8.4cm \epsfbox[ 0 0 571 421 ]
{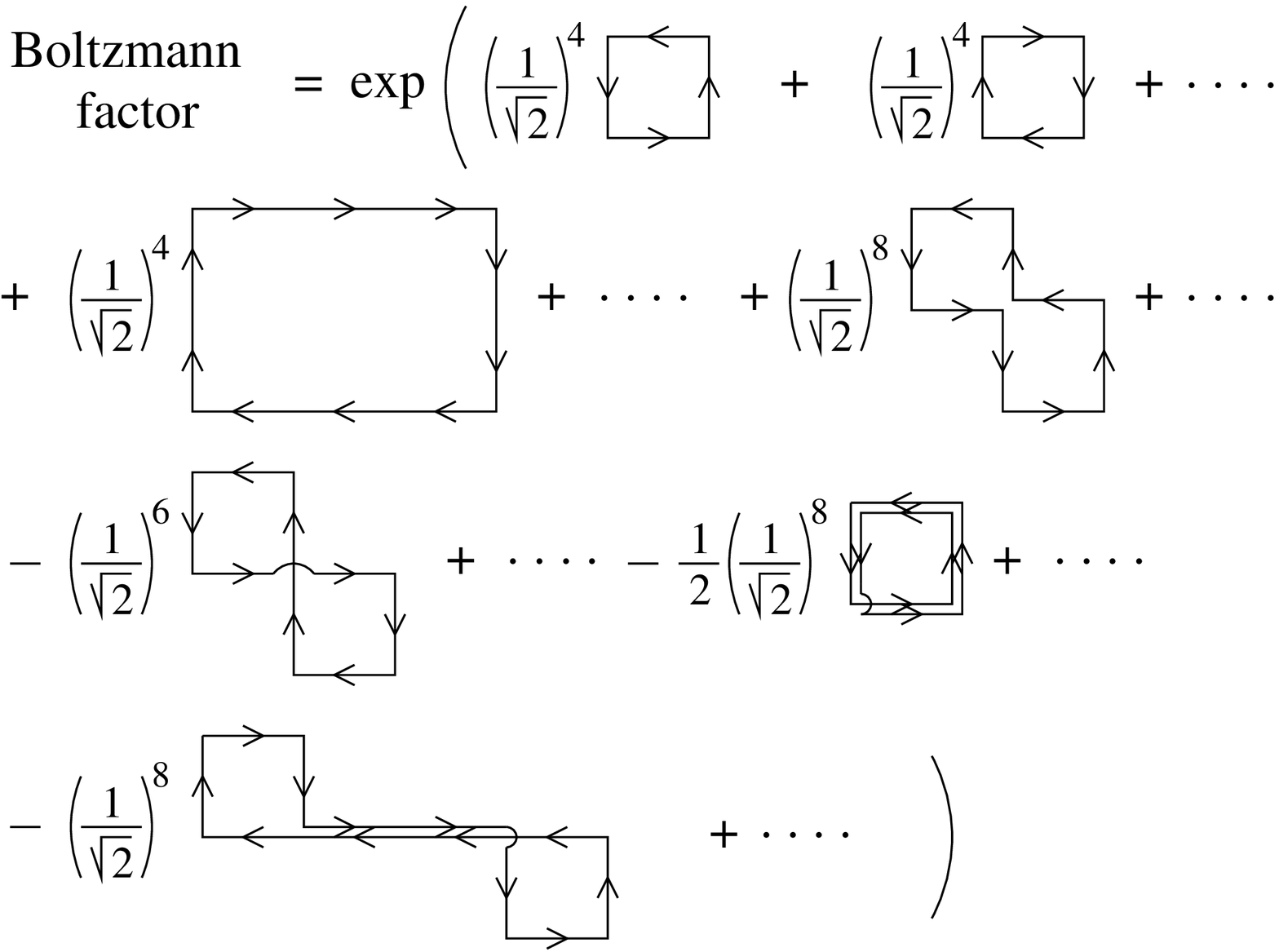}}
\caption{ {\sl Graphical representation for some of the terms in the
exponent of the Boltzmann factor. 
We display only the corner-, self-intersection- and iteration-factors.}
\label{exponentloops}}
\end{figure}

We approach the expansion by first discussing several examples which
will be used to extract the general algebraic rules for the
different terms along the way.  Keeping track of the terms in the
Taylor expansion of the
Boltzmann factor is most easily done by using the actual contours of
the loops associated with the different terms. So let's first look at
the different contributions to the sum in the exponent of the
Boltzmann factor.  In Fig.~\ref{exponentloops} we show some of the
terms as they appear in the exponent. 

The first two terms are two loops around a single plaquette and they
differ only in their orientation. Note that each of these two loops
(similarly for all loops shown) is
only a representative for all loops around a single plaquette, the
other loops differing by a translation, i.e. winding around any other
single plaquette of the lattice. These first two loops both have four
corners thus giving rise to a geometrical factor of
$(1/\sqrt{2})^4$. Certainly there are also the factors from taking the
product of the link variables around the plaquette. These are however
not relevant for the following discussion and thus are not displayed
explicitly. We show only the corner-, self-intersection- and 
iteration-factors.

Next we plot another rectangular loop, but now enclosing six
plaquettes. Again the corner-factor is $(1/\sqrt{2})^4$. The fourth
term is a loop enclosing two plaquettes but now having a pinch
structure. This loop has eight corners and no self-intersection such
that its geometrical weight  is $(1/\sqrt{2})^8$. The next term we
show is again a loop enclosing the same two plaquettes, the lower plaquette
however is now run through with opposite orientation. This loop 
has one  self-intersection, but only 6 corners, thus producing a
factor of $-(1/\sqrt{2})^6$.  Note that also the weight from the link
variables (not displayed in Fig.~\ref{exponentloops})
differs from the last contribution since  the lower
plaquette now has negative orientation and the link variables
for this plaquette are
complex conjugate. Next we show a loop which runs around a single
plaquette twice. It has eight corners and one self intersection. In
addition we have to take into account a factor of $1/2$ since the loop
iterates its contour twice. The total factor then is $- 1/2
(1/\sqrt{2})^8$. Finally we show another loop with a self-intersection. It 
consists of 2 loops around single plaquettes connected by a double line 
along two links. It has one self-intersection and 8 corners such that the
geometrical factor is $- (1/\sqrt{2})^8$.

Note that in Fig.~\ref{exponentloops} we display only a few terms
which are useful for the discussion of some examples below. It is
however straightforward to depict arbitrary terms from the series in
the exponent of (\ref{finalhop}) and to determine the corresponding
geometrical factors.  \\

Let's now perform the expansion of the Boltzmann factor in 
(\ref{finalhop}). We will
proceed as follows: In the  figures below we show on the left hand
side some configuration of links as it appears in the final 
representation, i.e.~after
the Boltzmann factor has been expanded. These contributions in their final
form will however be built from different loops and their products,
emerging when
expanding the Boltzmann factor. These different terms are shown in the
middle of the  figures below, together with their geometrical
weights. The total weight associated to the  graph depicted on the
left hand side is the sum of the weights for all contributing
terms. We give this number on the right hand side, separated by an arrow.
We remark that the expansion of the Boltzmann factor will produce 
contributions consisting of several disconnected pieces located on 
different positions of our lattice. Up to the combinatorial factors which
will be discussed below, disconnected pieces in different areas of the 
lattice do not communicate with each other. Our examples thus concentrate 
on the different terms locally building up a contribution. 
The inclusion of disconnected pieces located elsewhere is a trivial 
generalization.
\begin{figure}[htp]
\centerline{\hspace*{-2mm}
\epsfysize=1.18cm \epsfbox[ 0 0 402 59 ] {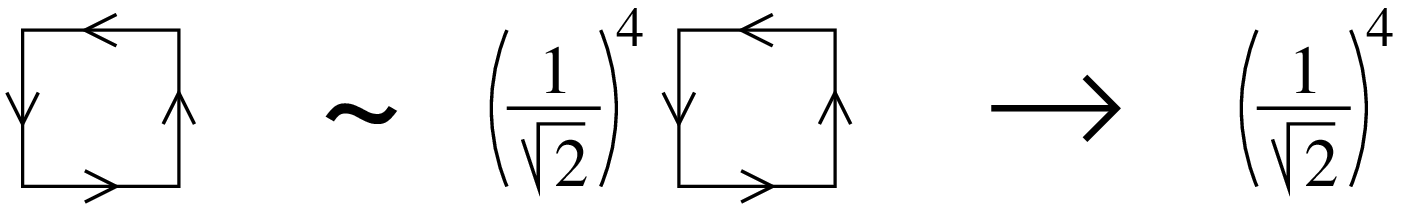}}
\caption{ {\sl A loop around a single plaquette.}
\label{example1}}
\end{figure}

In Fig.~\ref{example1} we show a loop which consists only of the four
links around a single plaquette. As is obvious from 
Eq.~(\ref{finalhop}) and its geometrical representation in
Fig.~\ref{exponentloops}, there is only one contribution in the Taylor
expansion of the Boltzmann factor giving rise to this contribution. It
is the first loop depicted in Fig.~\ref{exponentloops}. Furthermore
only the linear term in the power series for the exponential function
produces this term since for higher powers, this loop would
be multiplied with other loops. Thus we find only one term on the
right hand side of Fig.~\ref{example1}. The total weight for this term
is the same as we had in the exponent,  namely $(1/\sqrt{2})^4$. 
\begin{figure}[htp]
\centerline{
\epsfysize=1.3cm \epsfbox[ 0 0 599 66 ] {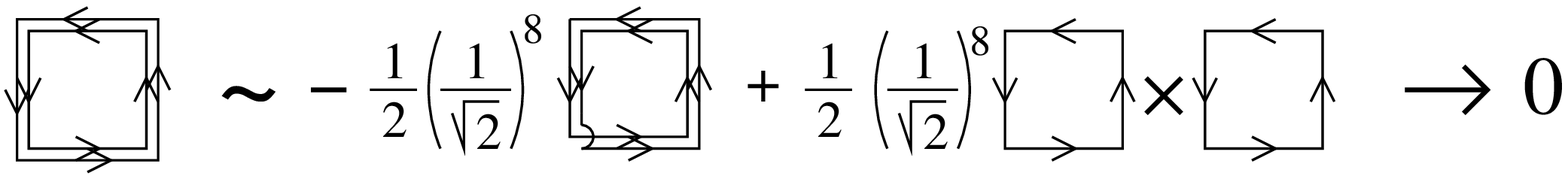}}
\caption{ {\sl A loop which iterates its contour twice.}
\label{example4}}
\end{figure}

Next we consider a contribution where a loop iterates it's contour 
twice (Fig.~\ref{example4}). In particular we consider the simplest 
such term, i.e.~the 
configuration where all links around a simple plaquette are
occupied twice. This configuration firstly can be resolved with a
single connected loop. This loop has eight corners and one
self-intersection which produces a minus sign. In addition we have to
take into account a factor of $1/2$ since this loop completely iterates
its contour twice. The overall factor thus is $- 1/2
(1/\sqrt{2})^8$. Secondly the configuration can be resolved as the
square of the loop which runs around our plaquette only once. 
Each of the single loops has four corners giving a total corner 
factor of $(1/\sqrt{2})^8$. In addition one has a factor 1/2 which is 
the factor for the quadratic term in the power series of the exponential
function. Putting things together we find that the two contributions
cancel each other. 

This type of cancellation which we have seen in the simple example of
Fig.~\ref{example4} is the manifestation of a general rule: 
Whenever a term contains a loop which iterates its complete 
contour twice the contribution gets canceled. The general argument 
goes as follows: Let $l_1$ be the loop which iterates its contour twice.
Assume, that it appears in a contribution which also contains $n-1$ other
loops $l_i, i = 2, ... n$, which are located somewhere else on the lattice. 
This contribution then appears in the $n$-th
power term of the Taylor series for the exponential function. 
The corresponding factor is: 
\begin{equation}
\frac{1}{n!} \frac{-1}{2} \Big(\frac{1}{\sqrt{2}}\Big)^{c(l_1)}
f(l_2) f(l_3) ... f(l_n) \; n! \; .
\label{iterfact1}
\end{equation}
The $1/n!$ factor on the left hand side comes from the power series
for the exponential function. 
Then follows a minus sign from the self-intersection of $l_1$ and a factor 
of $1/2$ since $l_1$ iterates its contour twice. Next we have the corner 
factor for $l_1$ followed by the factors $f(l_i)$ for the other loops
(corner, self-intersection- and iteration-factors). Finally we find a 
combinatorial factor of $n!$ which gives the number of different ways the 
$n$ terms can be arranged.
Since $l_1$ iterates its contour twice, the same contribution can also be
obtained by squaring a loop $l_1^\prime$ which has the same contour as 
$l_1$, but runs through it only once. This contribution then appears in the
$n+1$-th power term of the Taylor series for the exponential function and
comes with a factor of 
\begin{equation}
\frac{1}{(n+1)!} \Big(\frac{1}{\sqrt{2}}\Big)^{c(l_1)}
f(l_2) f(l_3) ... f(l_n) \; \frac{(n+1)!}{2} \; .
\label{iterfact2}
\end{equation}
Here the factor of $-1/2$ is gone, since we have no self-intersection and 
also the contour is not run through twice by a single loop. The corner 
factor stays the same since $c(l_1) = 2 c(l_1^\prime)$. Finally the 
combinatorial factor changes to $(n+1)!/2$, where the factorial gives the 
number of ways the $n+1$ terms can be arranged. However,  
$l_1^\prime$ appears twice, such that we have to divide by 2. 
Thus we find that (\ref{iterfact1}) is exactly the negative of 
(\ref{iterfact2}). This proves that whenever a loop iterates its contour 
twice it can be matched with a contribution which has the opposite sign
and gets canceled. 

In principle one can extend this proof and show via induction
that all contributions, where
a loop completely iterates its contour more than once, get canceled. There is
however a simpler argument. It is based on the fact,
that not more than two lines can flow
through a site. To understand this we have to remember that each occupied
link carries with it a factor of $1/M(x)$ as can be
seen from the last product in (\ref{finalhop}).  Thus when $n$ loops
run through a site $x_0$ they produce a factor of
$1/M(x_0)^n$. Combined with the overall factor of $M(x_0)^2$ we find
$M(x_0)^{2-n}$. When $n$ exceeds 2, i.e.~more than two loops run
through a site, this term is no longer polynomial. On the other hand
one finds by direct evaluation
that the expression for the determinant on the finite lattice 
is a polynomial in $M(x_0)$. Using the fact that for $m + \theta(x) > 0$
the series represents the determinant (compare the discussion above) we 
conclude, that all vertices where more than 2 lines run through a
site have to vanish. This implies that loops which iterate their contour 
more than twice cannot contribute and for the case of the doubly 
iterated loop we have 
given the explicit argument for its cancellation above, thus showing that 
eventually none of the iterating loops in the exponent contributes. 

Before we go back to the discussion of further examples, let's 
bring in another
harvest from the discussion in the last few paragraphs: The factors $1/n!$
from the Taylor series of the exponential function always get canceled and
the combinatorial factor simply is 1. This can be seen as follows:
For a contribution which is the product of $n$ loops one finds that only 
terms where $n$ different loops are multiplied can contribute (if a loop 
could appear more than once we would have loops which are multiply
 occupied, a case 
which we have just excluded). These products come with a combinatorial factor
of $n!$, which gives the number of ways the 
terms can be arranged. It cancels the factor of $1/n!$ 
from the Taylor series for the exponential function. 
Thus the algebraic factor is 
always equal to 1.
\\
\begin{figure}[htp]
\centerline{
\epsfysize=2cm \epsfbox[ 0 0 609 103 ] {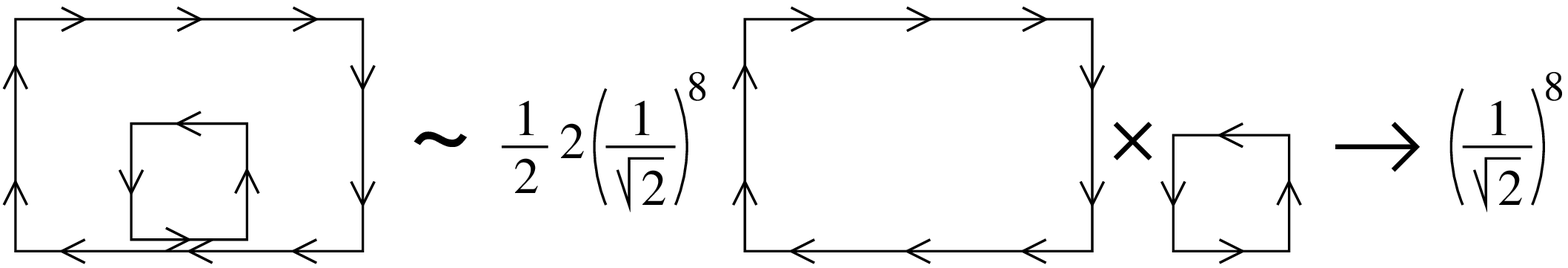}}
\caption{ {\sl Two nested loops with opposite orientation.}
\label{example2}}
\end{figure}

After this combinatorial excursion let's go back to discussing 
more examples. Next we analyze a graph with two nested loops
(Fig.~\ref{example2}). There is only one contribution in the
expansion of the Boltzmann factor giving rise to this graph. 
It is a product of two loops, i.e.~it comes from the quadratic
term in the Taylor series for the exponential function. Thus it has a
factor of $1/2$ which is the coefficient of the quadratic
term in the Taylor series. However, for our graph two different 
loops have to be
multiplied, making our contribution a cross term which comes 
with a factor of 2 canceling the factor $1/2$ from the power series. 
Each of the
two loops has 4 corners giving rise to a geometrical factor of
$(1/\sqrt{2})^8$. This is already the final factor associated with the
contribution depicted on the left hand side of Fig.~\ref{example2}.
\begin{figure}[htp]
\centerline{
\epsfysize=4.38cm \epsfbox[ 0 0 580 220 ] {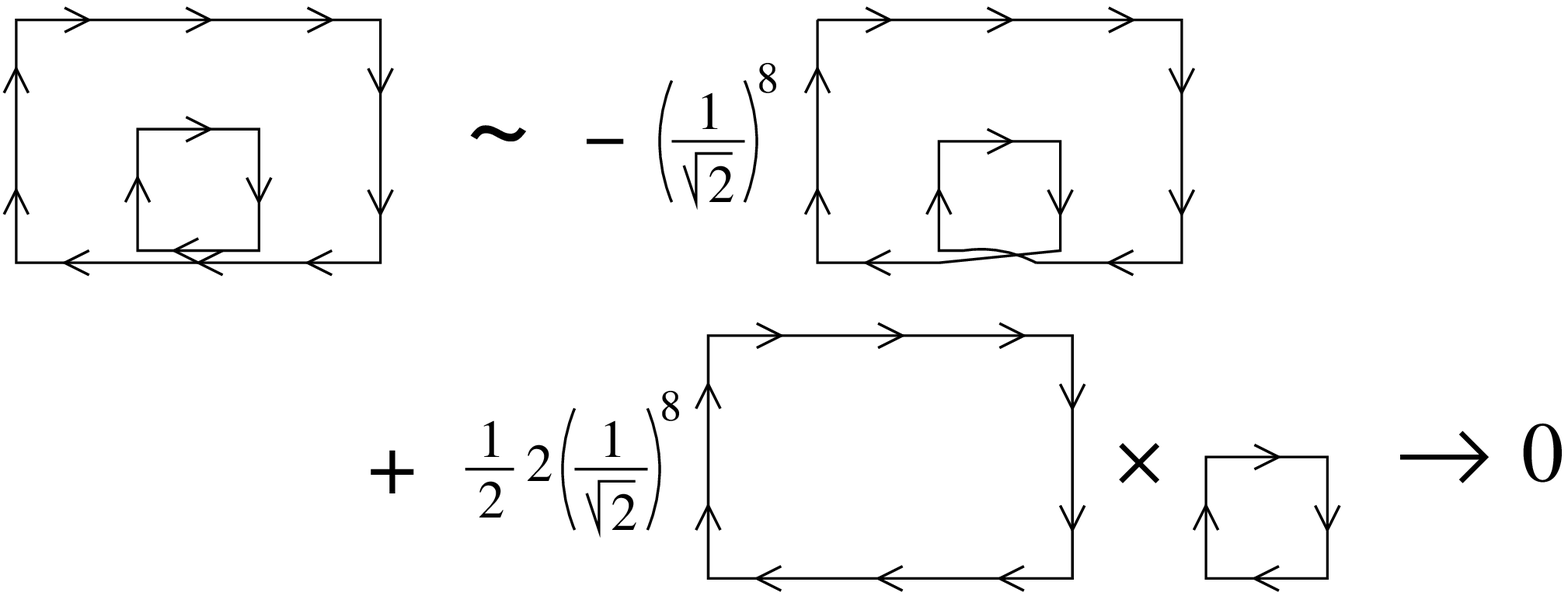}}
\caption{ {\sl Two nested loops with equal orientation.}
\label{example3}}
\end{figure}

In Fig.~\ref{example3} we look at a graph which contains the same
links, but the orientation of the inner loop has been reversed. In
addition to the contribution from a product of two loops, similar to
the last example, the new configuration also allows for a single
connected loop running through the chosen set of links. This term is
produced by the linear term of the power series for the exponential
function and thus simply has an algebraic factor of 1. The geometric
factor is $-(1/\sqrt{2})^8$, since the loop has eight corners and one
self-intersection. The product contribution has the same factor as in
the last example. The two contributions cancel each other.

Fig.~\ref{example3} demonstrates another general rule of our loop calculus:
The final expression will not contain any terms where a link is
occupied twice with the same orientation.  Whenever we compute the
weight for such a graph we find that the different terms in the
expansion of the Boltzmann factor cancel each other. In
Fig.~\ref{example3} this is demonstrated in an example, but let's
give the general argument: Whenever we have a configuration where
along  at least one link two lines run parallel with the same
orientation, then there are two ways to resolve this configuration
with terms coming from the expansion of the Boltzmann factor. One
contribution where the two lines do not cross and one with a
crossing. Both contributions have the same number of corners, the
latter one however obtains an extra minus sign due to the additional
self intersection. The two terms thus cancel each other. We remark
that at some other  parts of our loop there also might be structures
which can be resolved in different ways. However, already one link
which is run through twice in the same direction leads to a vanishing
of this contribution (compare also the discussion of completely iterated
loops given above). This exclusion of contributions with links
which are run through twice in the same direction can be interpreted as 
a manifestation of the Pauli principle  which forbids two excitations 
of the same state.
\begin{figure}[htp]
\centerline{
\epsfysize=2cm \epsfbox[ 0 0 632 103 ] {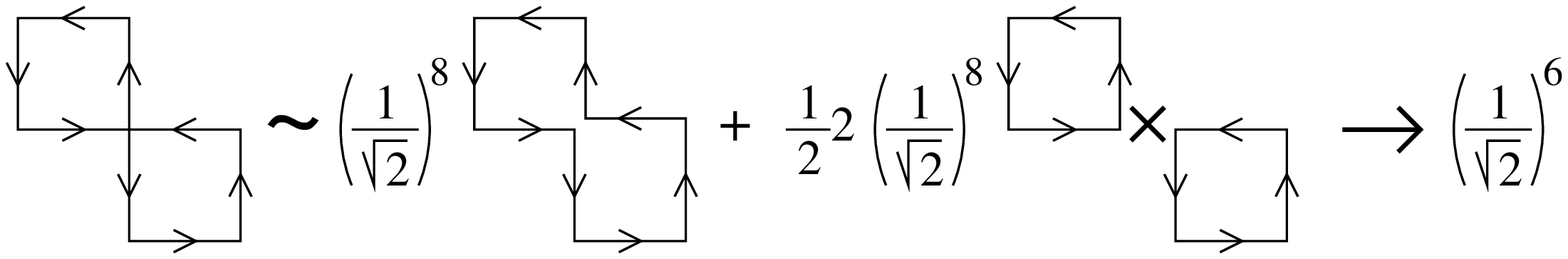}}
\caption{ {\sl A configuration with a self intersection.}
\label{example5}}
\end{figure}

Next we concentrate on how a configuration containing an intersection
can be resolved in different ways. 
The left hand side of Fig.~\ref{example5} shows a configuration which
contains two plaquettes with opposite orientation which share a common
corner. There is one possibility to run around the two plaquettes in a
single loop. This contribution comes from the linear term in the
Taylor series for the exponential function and has eight corners thus
acquiring a factor of $(1/\sqrt{2})^8$. Our two plaquettes can
also be obtained as the product of two loops, i.e.~a
contribution appearing in the quadratic term of the power
series. Again we have a total of eight corners giving a factor of
$(1/\sqrt{2})^8$ and when adding this to the coefficient of the linear
term we find an overall factor of $(1/\sqrt{2})^6$. 
\begin{figure}[htp]
\centerline{
\epsfysize=2cm \epsfbox[ 0 0 650 103 ] {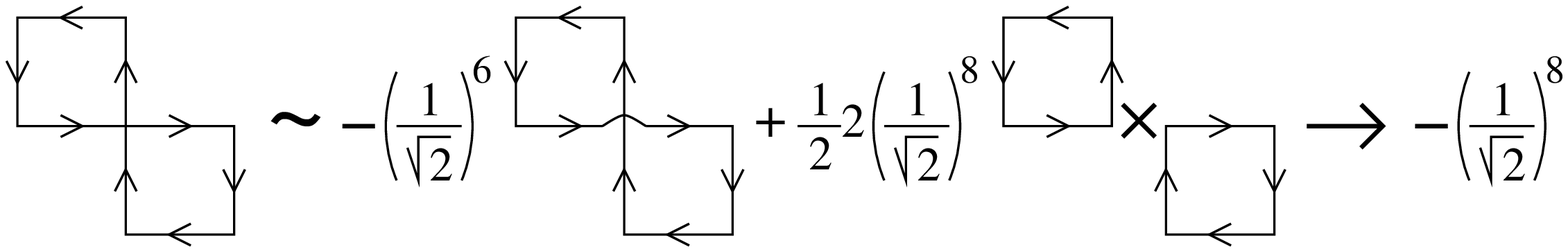}}
\caption{ {\sl Another configuration with a self
intersection.}
\label{example6}}
\end{figure}

In the configuration shown in Fig.~\ref{example6} we now have reverted
the orientation of the lower plaquette. Certainly this configuration
will still have a contribution from the quadratic term of the
exponential function where it is written as a product of two
loops. The geometrical factor is the same as in the last example. 
However, the linear
term, where our contour appears as a single connected loop, has
to be changed. It now has only six corners but one self-intersection
and thus a factor of $-(1/\sqrt{2})^6$. Adding this to the coefficient
of the product term we find an overall factor of $-(1/\sqrt{2})^8$.  
\begin{figure}[htp]
\centerline{
\epsfysize=2cm \epsfbox[ 0 0 604 102 ] {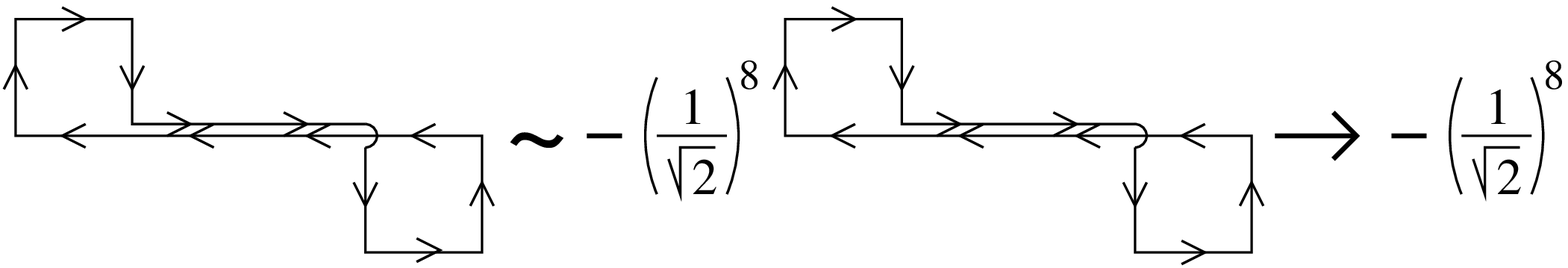}}
\caption{ {\sl Loops around two plaquettes joined by a double line.}
\label{example7}}
\end{figure}

Finally we discuss (Fig.~\ref{example7}) the contribution which corresponds 
to the last loop shown in Fig.~\ref{exponentloops}. It cannot be 
decomposed into the product of two loops and thus is entirely built 
from the linear term in the expansion of the exponential function.
The factor $-(1/\sqrt{2})^8$ is simply taken over from the exponent. 
Although not very interesting from a combinatorial point of view, this 
example demonstrates another important point. In order to avoid 
overcounting of self-intersections, we have to introduce a normal ordering
for double lines with two flows in opposite direction (the case of lines
with equal direction has already been ruled out above). We always draw a
line that flows from left to right above a line which flows from right 
to left. For vertical double lines we adopt the convention that downward 
flowing lines are drawn on the right hand side of upward flowing lines.

\begin{figure}[htp]
\centerline{
\epsfysize=16cm \epsfbox[ 0 0 502 633 ]
{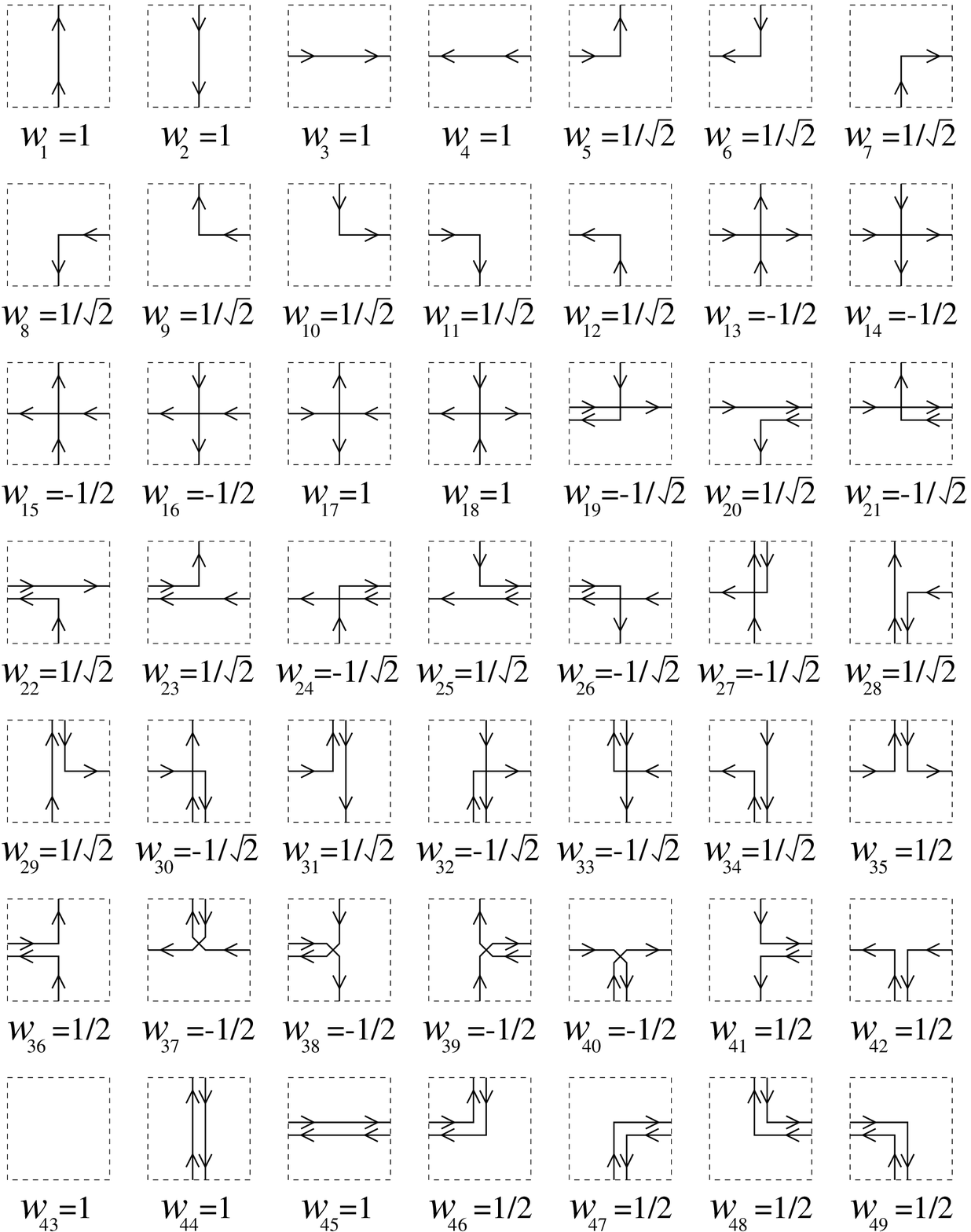}}
\caption{ {\sl The 49 tiles (vertices) with non-vanishing weights
$w_i$.}
\label{tiles}}
\end{figure}

After having analyzed several of these examples one finds that the
contributions on the left hand side can be decomposed into finitely
many elements and all contributions which appear in the expansion of
the Boltzmann factor can be built from these elements. We view these
elements as tiles with some configurations of oriented lines drawn on
these tiles (compare Fig.~\ref{tiles}). The tiles can now be arranged
on our rectangular lattice such that ingoing and outcoming flow lines
on neighboring tiles match each other. To each tile we assign a weight
$w_i$ and the total weight of a configuration is given by the product
of all weights for the tiles which were used in building the
configuration. When summing over all possible ways of different
tilings of our lattice (placing the centers of the tiles on the sites
of the lattice) we reproduce all possible fermion loops contributing
to the fermion determinant. The concept of writing partition functions
as a sum over tiles (vertices) was first introduced in
\cite{FaWu69,FaWu70} and these models are known as vertex models (see
also \cite{Ba82}).

The remaining problem is to assign the correct weights to our
tiles. From the example shown in Fig.~\ref{example1} we find that a
simple corner has to be assigned a factor of $1/\sqrt{2}$ which gives
the weights $w_5, ..., w_{12}$ in Fig.~\ref{tiles}. Fig.~\ref{example2}
shows that tiles with a straight line ($w_1, ..., w_4$ in
Fig.~\ref{tiles}) obtain a factor of 1. The same factor has to be
assigned when two straight lines run anti-parallel ($w_{44},
w_{45}$). The example discussed in Fig.~\ref{example7} furthermore shows
that antiparallel flow lines have to be normal ordered and our 
convention is to draw flow lines from left to right above flows from right 
to left and downward flows are draw right of upward lines.
From Fig.~\ref{example2} we learn, that tiles with a
straight line and a corner ($w_{19}, ..., w_{34}$ in Fig.~\ref{tiles}) have a
factor of $1/\sqrt{2}$ and an additional minus sign if the 
normal ordering of the double line enforces a crossing. 
Similarly one finds that tiles with two
corners  ($w_{35}, ..., w_{42}, w_{46}, ..., w_{49}$ in Fig.~\ref{tiles}) have
a factor of $(1/\sqrt{2})^2 = 1/2$, again with an additional minus sign if
a crossing occurs. 
Figs.~\ref{example4},\ref{example3} and the general argument given
there establish that all tiles which contain two lines running
parallel in the same direction must have vanishing weights 
(these tiles are not displayed in 
Fig.~\ref{tiles}).  From
Figs.~\ref{example5},\ref{example6} we find that for tiles showing
two crossing lines we have to assign a factor of  $-1/2$ when
the crossing can be resolved with a self-intersecting loop ($w_{13},
..., w_{16}$) and a factor of $1$ when no self-intersecting resolution is
possible ($w_{17}, w_{18}$). Finally we also have to include the empty tile
($w_{43}$) which has a factor of 1. In Fig.~\ref{tiles} we give a
complete list of all 49 tiles with non-vanishing weights.

We remark that tiles with an odd
number of incoming or outgoing lines of flow cannot appear, since all
contributions come from products of closed loops and thus the net flow
into a site has to vanish. Furthermore not more than two lines can 
run through a vertex as follows from the fact that the determinant 
is a polynomial in the local mass terms (see the discussion below 
Eq.~(\ref{iterfact2})).
The final expression for the fermion determinant reads
\begin{equation}
\det  K \; = \; \prod_{x \in \Lambda} M(x)^2 \; \sum_{t \in {\cal T}}
\; \prod_{j=1}^{49} (w_j)^{n_j(t)} \; \prod_{(y,\mu) \in t}
\frac{U_\mu(y)}{M(y)} \; .
\label{deterfinal}
\end{equation}
Here we introduced the set ${\cal T}$ of all admissible tilings of our
lattice $\Lambda$ with the tiles depicted in Fig.~\ref{tiles},
i.e.~the set of all arrangements of tiles such that the flow lines on
neighboring tiles match.  By $n_j(t)$ we denote the abundance of tile
number $j$ in a particular tiling $t \in {\cal T}$. The last product
runs over all links occupied by the tiling $t$. For links run through
in negative direction, the link variables have to be complex
conjugated. Our result is valid for two sets of boundary conditions:
1.) The fermions have open boundary conditions for both directions.
In this case also the vertex model has open boundary conditions for both
directions. 2.) The fermions have one direction with open boundary 
conditions and the second direction is compactified with anti-periodic 
boundary conditions for the fermions (cylinder geometry). The vertex model 
then has open boundary conditions for one direction, whereas the boundary 
conditions for the compact direction are periodic. 
\\

Let's now briefly discuss the applicability of our method to higher dimensional
theories and non-abelian gauge groups. The first step will again be a 
hopping expansion leading to formula (\ref{hopexp}). As in two dimensions this
series will converge and represent the determinant for $||H|| < 1$. Along the
lines of the proof in \cite{WeCh79} one can again establish this bound on 
the norm of the hopping matrix for a certain range of values for the mass $m$
and the auxiliary field $\theta$. Also the trace over the gamma matrices is 
known in four dimensions \cite{St81}. For the parameters where $||H|| < 1$ 
the Boltzmann factor can be expanded in the same way as in 2-d, and the
resulting series represents the determinant \cite{ReSi78}. Again we can apply
the power counting trick for the local mass terms and show, that the
number of fermion lines admissible at a single site is bounded. In four 
dimensions this bound is given by $4 N_c$ where $N_c$ denotes the number of 
colors. For non-abelian gauge fields in addition one has to take the trace
over the product of the link variables. The resulting formula is the loop 
representation of the fermion determinant. 

Such loop representations can be seen to lead to new ways for a numerical 
evaluation of the fermion determinant by directly updating the loop 
configurations \cite{GaGaSt99}. It is now 
possible to overcome the limitations of pseudo-fermion methods and simulate 
also odd numbers of flavors. For e.g~the one-flavor lattice Schwinger
model with Wilson fermions no numerical analysis exists and we are 
currently implementing algorithms based on the loop representation for 
this model \cite{GaGaSt99}. 

In this context it is interesting to note, 
that the loop representation for the fermion 
determinant also solves the fermion sign problem. This problem plagues
a direct evaluation of the Grassmann path integral \cite{Cr98}, as well
as the Hamiltonian treatment of fermionic systems \cite{ChWi99}. When
numerically updating fermions, the fermion sign problem enforces 
an exponentially increasing number of Monte-Carlo configurations as 
the volume of the system is increased or its temperature is lowered
\cite{ChWi99}. In our representation 
these permutation signs are gone and the remaining signs are due to the
traces over the gamma matrices. These latter signs are less severe,
since the weights for configurations with negative sign are exponentially
suppressed due to additional factors of $1/\sqrt{2}$ which are necessary 
to build loops with self intersections.  

Finally we argue that the loop representation could also lead to new
methods for the simulation of lattice field theories with non-vanishing
chemical potential. Although the Schwinger model with chemical potential is not
particularly interesting per se (see e.g.~\cite{AlNi98} for a study in the 
continuum) it nevertheless poses the same technical problems as QCD with
chemical potential. When introducing the chemical potential $\mu$ along the
lines of \cite{HaKa83} each link forward in time obtains a factor of
$e^{-\mu}$ and a factor of $e^{+\mu}$ for links backward in time. Thus
the symmetry between the two possible orientations of loops winding around 
the compact direction is broken and the determinant accquires a complex phase.
This complex phase prohibits the application of the pseudo-fermion method which
requires a non-negative kernel for the action of the pseudo-fermions. The loop
representation allows for an update of single loops and the chemical 
potential can be included in a natural way independent of the number of 
dimensions. Also this idea is currently being tested \cite{GaGaSt99} using
Thirring- and Gross-Neveu-type models in loop representation.

\section{Two limiting cases of our representation}
\noindent
Our final representation (\ref{deterfinal}) of the fermion determinant
has a relatively simple form. It is however instructive to reduce it
to limiting cases and compare it to known results for these instances
which were obtained with different methods. In particular we will
discuss the case of the determinant in a scalar background field 
and the case of the strongly coupled Schwinger model.  
\\   

We start with the case of the reduction to the case of the fermion
determinant in a scalar background field \cite{Ga98a,Ga98b}. There the result
for the fermion determinant reads
\begin{equation}
\det  K \; = \; \prod_{x \in \Lambda} M(x)^2 \;
\left( \sum_{l \in {\cal L}_{sa}} \Big(\frac{1}{\sqrt{2}}\Big)^{c(l)}
\prod_{y \in l} \frac{1}{M(y)} \right)^2 \; .
\label{deterscalar}
\end{equation}
The sum now runs over the set ${\cal L}_{sa}$ of self-avoiding, closed,
non back-tracking loops. Self-avoiding means that the loops may neither 
self-intersect nor touch each other, they can, however, consist 
of several disconnected pieces. The loops in ${\cal L}_{sa}$ are 
not oriented, i.e.~each loop is counted only with one of its two possible 
orientations. $c(l)$ again denotes the number of corners of $l$ and the 
product in (\ref{deterscalar}) runs over all sites $y$ visited by the
loops. 

Finding a general proof which directly shows the equivalence of the right 
hand side of (\ref{deterfinal}) at $U_{\mu}(x) = 1$ with the right hand side
of (\ref{deterscalar}) turns out to be a rather hard combinatorial problem. 
Instead we here discuss an example which illustrates how the two 
representations are related. The main difference is that the representation
(\ref{deterfinal}) for the determinant where also a gauge field is coupled 
has to work with oriented loops, since the link variables allow
to distinguish between forward and backward hopping (complex conjugation 
in the kernel (\ref{kernel})). Loops in the determinant for a scalar 
background field give the same contribution when their orientation is 
reverted, and thus it is possible to find the representation 
(\ref{deterscalar}) in terms of non-oriented loops. 
\begin{figure}[ht]
\centerline{
\epsfysize=2.8cm \epsfbox[ 0 0 517 143 ] {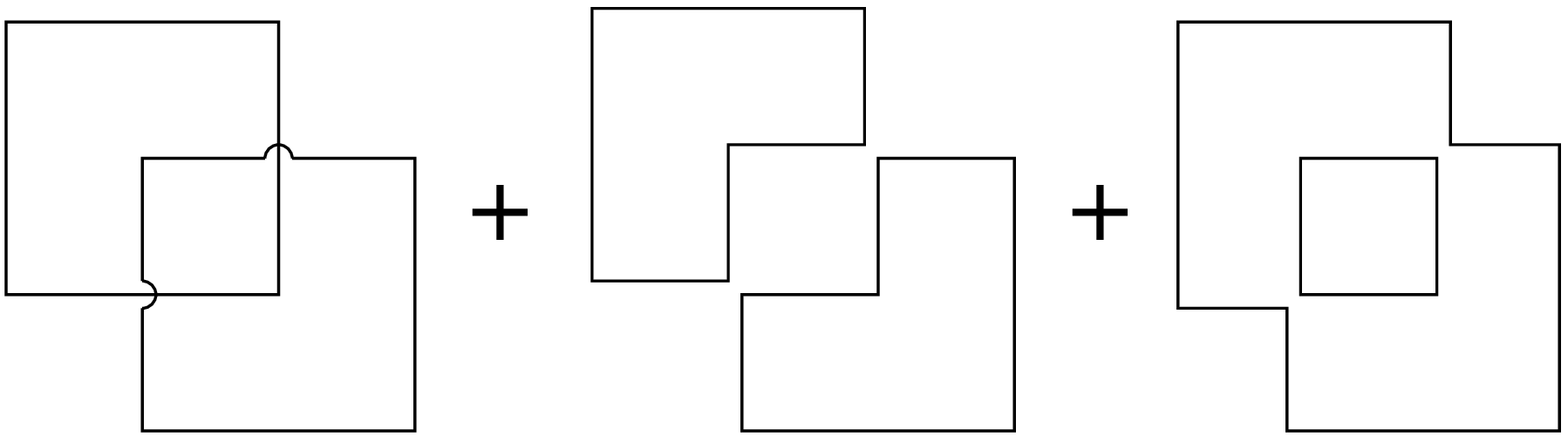}}
\caption{ {\sl The contributions to a given configuration as they
appear in the scalar formula (\protect{\ref{deterscalar}}).}
\label{equiv1}}
\end{figure}
\begin{figure}[ht]
\centerline{
\epsfysize=3.8cm \epsfbox[ 0 0 535 191 ] {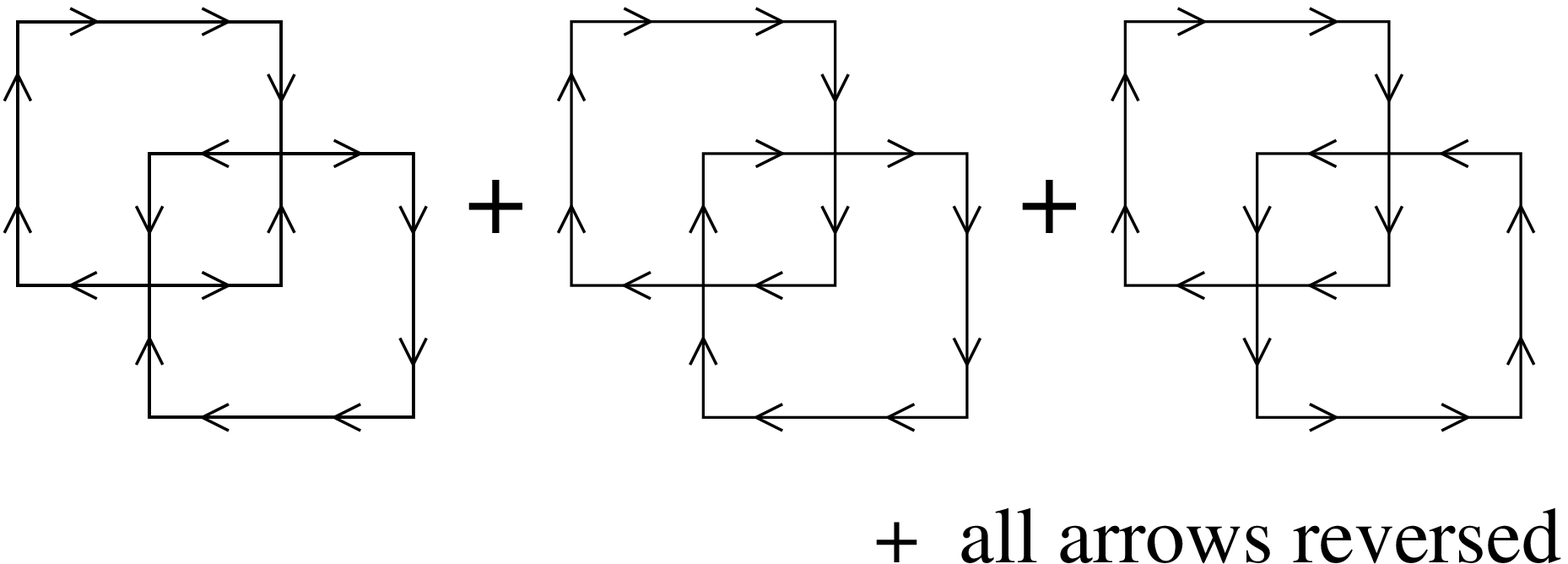}}
\caption{ {\sl The same contribution as in 
Fig.~\protect{\ref{equiv1}}, now being built from the oriented loops of 
formula (\protect{\ref{deterfinal}}).}
\label{equiv2}}
\end{figure}

Let's now look at a simple configuration which we depict in 
Fig.~\ref{equiv1}. In the figure we show how this term is generated in 
formula (\ref{deterscalar}), i.e.~how it can be written as the 
product of two self-avoiding loops. There are three ways of doing so,
each of them is a cross term in the square, such that we collect an overall 
factor of 2. The first term gives rise to 8 corners, while the
other two configurations each have a total of 12 corners. We thus find
a factor of (the products of $1/M(y)$ are irrelevant here and we suppress
them)
\[
2 \left[ \Big( \frac{1}{\sqrt{2}} \Big)^8 + 
\Big( \frac{1}{\sqrt{2}} \Big)^{12} + \Big( \frac{1}{\sqrt{2}} \Big)^{12} 
\right] \; = \; \frac{3}{16} \; .
\]
Next we generate the same term, now using the tiling formula 
(\ref{deterfinal}). Again we find 3 topologically different configurations  
which we show in Fig.~\ref{equiv2}. In addition these three configurations 
come in pairs related to the plotted terms by reverting all arrows. This 
gives again an overall factor of 2. By collecting the factors for the
vertices as given in Fig.~\ref{tiles} we find
\[
2 \left[  \Big( \frac{1}{\sqrt{2}} \Big)^8 \times 1 \times 1 
+ \Big( \frac{1}{\sqrt{2}} \Big)^8 \times \frac{1}{2} \times \frac{1}{2}
+ \Big( \frac{1}{\sqrt{2}} \Big)^8 \times \frac{1}{2} \times \frac{1}{2}
\right] \; = \; \frac{3}{16} \; .
\]
Thus both formulas (\ref{deterfinal}) and (\ref{deterscalar}) give the 
same factor. Similarly one can show the equivalence of (\ref{deterfinal}) and
(\ref{deterscalar}) for more complicated examples.
\\

\begin{figure}[htp]
\centerline{
\epsfysize=1.7cm \epsfbox[ 0 0 684 101 ] {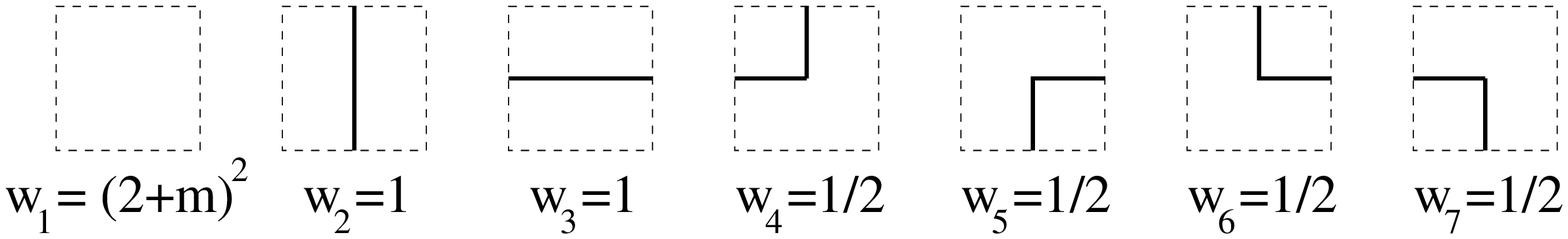}}
\caption{ {\sl The 7 vertices occuring in the strong coupling limit
and their weights.}
\label{7vertex}}
\end{figure}

Let's now proceed to the case of the strongly coupled Schwinger model.  All
scalar field variables are set to zero, $\theta(x) = 0$, and  we write
the U(1) link variables now explicitly as
\[
U_\mu(x) \; = \; e^{i A_\mu(x)} \; \; \; \; , \; \; \; \; A_\mu(x) \in
(-\pi, +\pi ] \; .
\]
The partition function of the lattice Schwinger model is then given by 
\begin{equation}
Z_g \; = \; \int_{-\pi}^\pi \prod_{(x,\mu)} \frac{d A_\mu(x)}{2 \pi}
\; e^{-\frac{1}{g^2} S_{gauge}[A]} \;  \det K \; .
\end{equation}
At strong (= infinite) coupling $g$, the Boltzmann factor for the
gauge field equals 1.  Only the determinant appears in the integrand.
Each $t$ in the sum over  all tilings ${\cal T}$
gives rise to links being empty, singly occupied links and doubly
occupied links with link variables complex conjugate to each
other. These cases correspond to the integrals
\[
\int_{-\pi}^\pi \frac{d A_\mu(x)}{2\pi} = 1 \; , \;  \int_{-\pi}^\pi
\frac{d A_\mu(x)}{2\pi} e^{\pm i A_\mu(x)} = 0 \; , \; \int_{-\pi}^\pi
\frac{d A_\mu(x)}{2\pi} e^{+ i A_\mu(x) - i A_\mu(x)} = 1 .
\]
The second integral implies
that whenever a configuration of tiles contains a link which
is occupied only once, the whole configuration vanishes when
integrating the gauge field. Thus only configurations  with either
empty or doubly occupied links (with opposite orientation)
contribute in the sum over all
tilings, i.e.~only the tiles $43, ..., 49$ in Fig.~\ref{tiles}
contribute. The strong coupling limit thus reduces the lattice
Schwinger model to a seven vertex model with vertices and weights as
depicted in Fig.~\ref{7vertex}, where we have included the overall
factor $\prod_x (2 + m)^2$ into the monomer weight $w_1$. Note that 
after integrating the gauge fields in the strong coupling limit 
the orientation of the link variables gets lost. Our loops do no longer
have an orientation and it is sufficient to have single lines on the tiles
in Fig.~\ref{7vertex}.

The partition function in the strong coupling limit reads
\begin{equation}
Z_{g = \infty} \; = \; \sum_{t \in {\cal T}_7} \; \prod_{j=1}^7
(w_j)^{n_j(t)} \; .
\label{7vertexz}
\end{equation}
Here ${\cal T}_7$ denotes the set of admissible tilings obtained from 
the seven tiles depicted in Fig.~\ref{7vertex}.
Formula (\ref{7vertexz})
is equivalent to the partition function for
self-avoiding loops with corner activity 1/2 and monomer weight
$(2+m)^2$.
The representation (\ref{7vertexz}) for the strongly coupled Schwinger
model was first found by  Salmhofer \cite{Sa91}. Since \cite{Sa91}
uses different techniques (expansion of the Boltzmann factor and local
saturation of the Grassmann integral) the derivation of this result
as a limiting case of our  
general formula provides a second non-trivial test of
Eq.~(\ref{deterfinal}).

It is interesting to note that our representation (\ref{deterfinal}) 
allows to generalize the vertex model representation for the strongly
coupled case to a model with $N$ flavors. For $N$ flavors the integration 
of the fermions gives the 
$N$-th power of the fermion determinant. This simply corresponds to 
superimposing $N$ different 49-vertex models with loops in $N$ different 
colors onto each other. Integrating the gauge fields in the strong coupling 
limit discards all vertices where at least one link is only singly 
occupied. For the case of the strongly coupled two-flavor Schwinger model 
this leads to a 163-vertex model \cite{cbl99}.

We remark that the representation of the strongly coupled lattice Schwin\-ger 
model as a seven vertex model demonstrates the power of loop representations
when performing numerical simulations. The model (\ref{7vertexz}) has been
studied numerically in \cite{GaLaSa92,KaMeTu95} and high precision data on 
the phase
diagram has been obtained. For vertex models furthermore the use of 
cluster algorithms might become possible which could further increase the
quality of numerical data \cite{EvLaMa93}. 
\\

\section{Integrating out the gauge fields at arbitrary coupling}
\noindent
We now integrate the gauge fields at arbitrary coupling for the model
with open boundary conditions. The basic idea is to fill the loops with
plaquette variables. We remark that when choosing our lattice to be the
above discussed cylinder we find topologically non-trivial loops. For these
loops the gauge field integral, although still solvable in closed form,
is more involved and for convenience we here discuss only the case of 
a rectangular lattice with open boundary conditions.

\begin{figure}[htp]
\centerline{
\epsfysize=4.75cm \epsfbox[ 0 0 507 237 ] {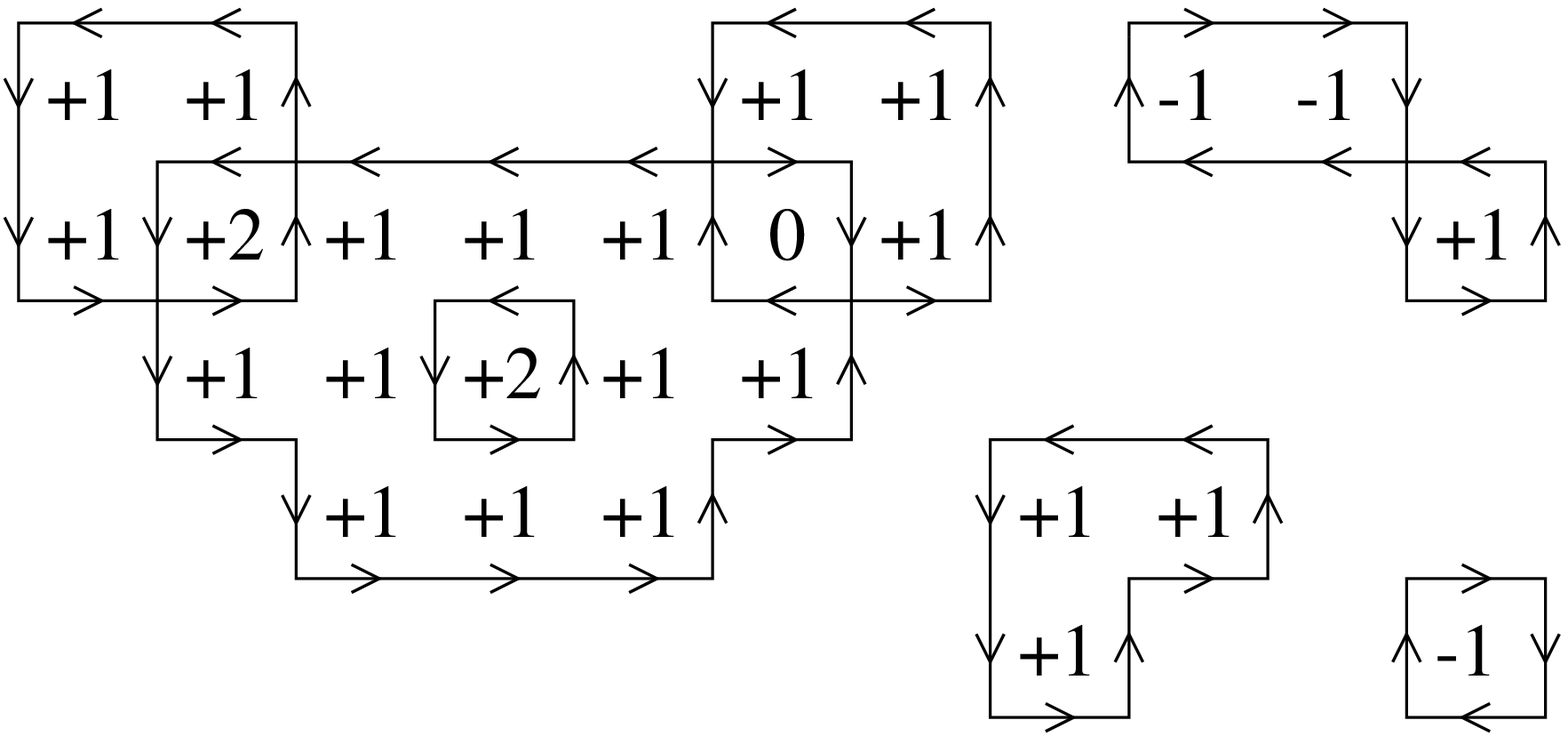}}
\caption{ {\sl A contribution to the determinant and the occupation numbers 
for the plaquettes inside the loops.}
\label{occupation}}
\end{figure}

We now assign plaquette variables $V(p)$ to each plaquette $p$, 
which are given 
by the product of the 4 link variables associated to the plaquette $p$  when 
running through it in mathematically positive direction,
\begin{equation}
V(p) \; = \; \prod_{(x,\mu) \in p} U_\mu(x) \; .
\label{plaqvar}
\end{equation}
To each plaquette $p$ we furthermore assign an occupation number $J_p$ 
which gives the power of $V(p)$ for the plaquette $p$. Negative $J_p$ 
is used when we assign the complex conjugate plaquette variable $V(p)^*$
with exponent $|J_p|$. When two neighboring plaquettes have the same 
occupation numbers, the link variable shared by the two plaquettes
cancels. When the occupation numbers differ, the shared link variable 
remains. Note that for neighboring plaquettes the occupation numbers can 
only differ by $\pm 1$, since we can never have more than one unit of flux
on a link. Fig.~\ref{occupation} gives an example of a configuration 
of loops and how its occupation numbers are assigned.

For a given configuration of occupation numbers $J_p$ the gauge field 
integral reads,
\begin{equation}
I_g[J_p] \; = \; \int \prod_{(x,\mu)} d U_\mu(x) \; 
e^{-\frac{1}{g^2} S_{gauge} } \; \prod_p \left( \prod_{(y,\nu) \in p}
U_\nu(y) \right)^{J_p} \; .
\end{equation}
In our two-dimensional model this integral can easily be solved by choosing 
a simple gauge, i.e.~a maximal tree of link variables set to 1, which 
factorizes the gauge field integral. The gauge field integral is simply
a product over all plaquettes with the factors determined by the occupation
numbers,
\begin{equation}
I_g[J_p] \; = \; \prod_p F_g(J_p) \; ,
\label{gaugeintegral}
\end{equation}
where the function $F_g(n), n \in$ Z\hspace{-1.3mm}Z is given by (we quote
the results for Wilson action
as well as for Villain action),
\begin{equation}
F_g(n) \; = \; \left\{ \begin{array}{ll}
e^{-\frac{1}{g^2}} \; \; 
\mbox{I}_{|n|}\Big(\frac{1}{g^2}\Big) \; \; \; &   
\mbox{(Wilson action)} \; \; ,  \\
 & \\
\frac{g}{\sqrt{2 \pi}} \; e^{-\frac{g^2}{2} n^2} \; \; \; &  
\mbox{(Villain action)} \; \; . \end{array} \right. 
\label{plaquettefactors}
\end{equation}
Here I$_n$ denotes the modified Bessel function. 
For the partition function of the lattice Schwinger model we thus find,
\begin{equation}
Z_g \; = \; \int \prod_{(x,\mu)} dU_\mu(x)\; 
e^{-\frac{1}{g^2} S_{gauge}} \det K
\; = \; \sum_{t \in {\cal T}}
\; \prod_{j=1}^{49} (\tilde{w}_j)^{n_j(t)} \; 
\prod_p F_g(J_p(t)) \; .
\label{schwingerfinal}
\end{equation}
The last term denotes the product over all plaquette factors depending 
on the occupation numbers $J_p(t)$ for a given tiling $t$. We remark that we 
now have included the factors of $(2 + m)$ into the new weights $\tilde{w}_j$.
They are obtained by multiplying the old weight $w_{43}$ by $(2+m)^2$, 
by multiplying weights $w_{1}, ... , w_{12}$ with $2 + m$ and setting
$\tilde{w}_j = w_j$ for the remaining weights. 

Our expression (\ref{schwingerfinal}) represents the partition 
function as 
a sum over discrete variables. This expression
could lead to the possibility of 
applying cluster algorithms \cite{EvLaMa93} which might considerably improve 
the quality of numerical data. 

\section{Observables}
In this section we discuss the loop representations for 
pure gauge field observables 
such as the Wilson loop and for local fermionic observables in particular
scalars and vectors.
\\

Let's start with the simpler observables which are the ones only 
containing gauge fields.
The prototype of such an observable is the Wilson loop
\[
W_{\cal C}(U) \; = \; \prod_{(x,\mu) \in {\cal C}} U_\mu(x) \; .
\]
Here ${\cal C}$ is a closed contour and without restriction of generality 
we assume that it is oriented in mathematically positive direction. As for 
the fermion loops, we can create the link variables on ${\cal C}$ by filling
the whole contour with plaquette variables as defined in (\ref{plaqvar}). 
This requires an occupation number of $+1$ for all plaquettes inside 
${\cal C}$. This occupation number can simply be added to the ocupation 
numbers $J_p$ from the fermion loops. When integrating out the gauge fields
we find for the vacuum expectation value of the Wilson loop in the lattice
Schwinger model (compare (\ref{schwingerfinal})) 
\begin{equation}
\left\langle W_{\cal C}(U) \right\rangle_{g} \; = \; \frac{1}{Z_g} 
 \sum_{t \in {\cal T}}
\; \prod_{j=1}^{49} (\tilde{w}_j)^{n_j(t)} \; 
\prod_p F_g \Big( J_p(t) + \chi_p({\cal C}) \Big) \; ,
\end{equation}
where we defined $\chi_p({\cal C})$ to be the characteristic function of 
the area inside the chosen contour ${\cal C}$ 
with $\chi_p({\cal C}) = 1$ for plaquettes $p$ 
in the interior of ${\cal C}$ and $\chi_p({\cal C}) = 0$ for plaquettes 
outside. It is obvious that this formula can easily be generalized to 
more involved gauge invariant observables for the gauge field. 
\\

Only slightly more complicated is the loop representation for local 
fermi\-onic 
observables, in particular scalars and vectors. Here we can make use of 
the fact, that our formalism gives the fermion determinant, i.e.~the 
fermionic path integral, in an arbitrary gauge field and in an arbitrary 
local 
scalar field $\theta(x)$. Thus by differentiating the determinant with respect 
to these external fields we can generate n-point functions for the fermionic
bilinears in the action, i.e.~scalars and vectors. By writing the link 
variables as $U_\mu(x) = \exp(i A_\mu(x))$ with $A_{-\mu}(x) = 
-A_\mu(x - \hat{\mu})$, the kernel $K$ for the fermion action 
reads (compare (\ref{kernel}))
\[
K(x,y) \; = \; \Big[2 + m + \theta(x)\Big] \; \delta_{x,y}  \; - \;
\sum_{\mu = \pm 1}^{\pm 2} \frac{ 1 \mp \sigma_\mu}{2} \;
e^{ i A_\mu(x)} \;
\delta_{x+\hat{\mu},y} \; .
\]
Thus we obtain for the expectation value (just the fermion integration)
of scalars and 
vectors
\begin{eqnarray}
& & \left\langle \prod_{k = 1}^M \overline{\psi} \psi(v_k) 
\prod_{l=1}^N j_{\nu_l}(z_l) \right\rangle_{fermions}  
\nonumber \\
& = &\int d\overline{\psi} d \psi \; e^{-\sum_{x,y} 
\overline{\psi}(x) K(x,y)\psi(y)} \; 
\prod_{k = 1}^M \overline{\psi} \psi(v_k) \; 
\prod_{l=1}^N j_{\nu_l}(z_l) \nonumber \\
& = & (-1)^M \prod_{k = 1}^M \frac{\partial}{\partial \theta(v_k)} \; 
(i)^N \prod_{l=1}^N \frac{\partial}{\partial A_{\nu_l}(z_l)}
\int d\overline{\psi} d \psi \; 
e^{-\sum_{x,y} \overline{\psi}(x) K(x,y)\psi(y)}
\nonumber \\
& = & (-1)^M \prod_{k = 1}^M \frac{\partial}{\partial \theta(v_k)} \; 
(i)^N \prod_{l=1}^N \frac{\partial}{\partial A_{\nu_l}(z_l)}
\; \det K \; .
\nonumber
\end{eqnarray}
The current $j_\nu(z)$ is the Noether current corresponding to the gauge 
symmetry of our action 
\begin{eqnarray}
j_\nu(z) & = & \frac{1}{2} 
\Big[ \overline{\psi}(z) \sigma_\nu U_\nu(z) \psi(z+\hat{\nu})  
+ \overline{\psi}(z+\hat{\nu}) \sigma_\nu U_\nu(z)^* \psi(z) \Big] 
\nonumber \\ 
& - &
\frac{1}{2} 
\Big[ \overline{\psi}(z) U_\nu(z) \psi(z+\hat{\nu})  
- \overline{\psi}(z+\hat{\nu}) U_\nu(z)^* \psi(z) \Big] \; .
\end{eqnarray}
In the continuum limit the first term turns into the vector-current 
$\overline{\psi}(z) \gamma_\nu \psi(z)$ while the second term is of 
order ${\cal O}(a)$ and vanishes.

Now we act with the differential operators $\partial/\partial \theta(v)$
and $\partial/\partial A_\nu(z)$ on our loop expression 
(\ref{deterfinal}). When making explicit all external fields the formula
(\ref{deterfinal})
reads (now again using the original vertex weights $w_i$ from Fig.~\ref{tiles}
and not the rescaled vertices $\tilde{w}_i$ used in (\ref{schwingerfinal}))
\begin{equation}
\det  K \; = \; \prod_{x \in \Lambda} [2 + m + \theta(x)]^2 \; 
\sum_{t \in {\cal T}}
\; \prod_{j=1}^{49} (w_j)^{n_j(t)} \; \prod_{(y,\mu) \in t}
\frac{e^{\pm i A_\mu(y)}}{2 + m + \theta(y)} \; .
\label{explicitfields}
\end{equation}
In the 
numerator of the last term which collects the link factors, the plus sign 
in the exponent has to be taken for links $(y,\mu)$ which are run through 
in positive direction for the configuration $t$  
and the minus sign is taken for links occupied in negative 
direction. When acting with $\partial/\partial \theta(v)$ for some fixed 
site $v$ on the expression (\ref{explicitfields}) we have to distinguish three
different cases. When two fermion lines run through $v$, all 
terms containing $\theta(v)$ cancel each other and the differentiation 
simply gives zero. When one fermion line runs through $v$ this contribution 
is linear in $2 + m + \theta(v)$ and after differentiation there 
remains a factor of 1. Finally when the site $v$ is empty the factor 
in (\ref{explicitfields}) is    
$[2 + m + \theta(v)]^2$ and the action of the differential operator then
gives a 
weight of $2[2 + m + \theta(v)]$. The results for these three cases can be 
denoted conveniently by inserting the following factor in the sum over 
the configurations $t \in {\cal T}$ (we have now set the scalar field 
$\theta$ to zero everywhere which gives us the expression for the Schwinger 
model)
\begin{equation}
\Big[ 1 - S_v^{(2)}(t) \Big] \left( \frac{1}{2+m} \right)^{S_v^{(1)}(t)}
\left( \frac{2}{2+m} \right)^{S_v^{(0)}(t)} \; ,
\label{scalarfactor}
\end{equation}
where we defined $S_v^{(n)}(t)$ to be equal to 1 when the site $v$ is visited 
exactly $n$-times by the fermion loops of the configuration $t$ and zero 
otherwise. The first factor in (\ref{scalarfactor}) produces a zero whenever 
2 fermion lines run through $v$. The other two terms correct the 
weight at $v$ for 
the cases of a single fermion line and for $v$ being empty.

The result for the differential operator $i \partial / \partial A_\nu(z)$ is
even simpler to denote. From (\ref{explicitfields}) it is clear that for 
configurations $t$ where the link $(z,\nu)$ is empty, or configurations 
where this link is doubly occupied and the link variables cancel each other 
due to their opposite orientation, this contribution does not depend on 
$A_\nu(z)$ and is annihilated by $i \partial / \partial A_\nu(z)$. For 
configurations where the link $(z,\nu)$ is run through with positive 
orientation, we obtain after applying $i \partial / \partial A_\nu(z)$
a factor of $-1$ and a factor of $+1$ for negative orientation. Thus the 
differential operator simply measures the negative of the net flow of
fermion lines through the link $(z,\nu)$. By defining
\begin{equation}
L_{(z,\nu)}(t) \; = \; \left\{ \begin{array}{cl}
+1  & \mbox{when $t$ occupies $(z,\nu)$ in positive direction} \\
-1  & \mbox{when $t$ occupies $(z,\nu)$ in negative direction} \\
0   & \mbox{when $t$ has net flux zero through $(z,\nu)$}
\end{array} \right. \; ,
\label{vectorfactor}
\end{equation} 
we can take into account the vectors by simply including $-L_{(z,\nu)}(t)$
in the sum. Inserting this factor, together with (\ref{scalarfactor}) into
the sum over all configurations we obtain for the fermionic n-point
functions in the lattice Schwinger model (the gauge fields 
were integrated out as above, $\theta(x)$ is set to 0, and we use again 
the rescaled weights $\tilde{w}_i$ as defined below equation 
(\ref{schwingerfinal}))
\begin{eqnarray}
& & \hskip-8mm \left\langle \prod_{k = 1}^M \overline{\psi} \psi(v_k) 
\prod_{l=1}^N j_{\nu_l}(z_l) \right\rangle_{g} =  
(-1)^{M+N} \frac{1}{Z_g} \sum_{t \in {\cal T}}
\; \prod_{j=1}^{49} (\tilde{w}_j)^{n_j(t)} \; 
\prod_p F_g(J_p(t)) 
\nonumber \\
& \times & 
\prod_{k = 1}^M \left[ \Big[ 1 - S_{v_k}^{(2)}(t)\Big] \left( 
\frac{1}{2+m} \right)^{S_{v_k}^{(1)}(t)}
\left( \frac{2}{2+m} \right)^{S_{v_k}^{(0)}(t)} \right]
\prod_{l=1}^N L_{(z_l,\nu_l)}(t) \; .
\label{fermnpoint}
\end{eqnarray}
After getting used to it, this expression turns out to be very powerful and 
also very simple from a numerical point of view. For computing n-point 
functions of scalars one merely has to locally change the weights on the sites
supporting the scalars, and for vectors only the net flux 
through the links associated with the currents has to be counted. 
This is certainly much simpler
than inverting the fermion matrix as is necessary in the standard 
numerical approach.

It is instructive to play a little bit with formula (\ref{fermnpoint}), so 
let's e.g.~show that $\langle j_\nu(z)\rangle_{g} = 0$. As discussed above 
only configurations where the link $(z,\nu)$ is singly occupied can 
contribute. These configurations, however, come in pairs related by reversing
the orientation of all fermion lines. The two contributions have the
same weight from the gauge fields ($J_p \rightarrow - J_p$) but are counted 
with opposite sign in (\ref{fermnpoint}) and thus the one-point function of
the current vanishes as it should. 

Furthermore it is interesting to analyze some variables in the strong coupling
limit, in particular the chiral condensate 
$\langle \overline{\psi} \psi (v) \rangle_{g=\infty}$. 
As discussed above, in the strong
coupling limit all configurations with singly occupied links are suppressed. 
Thus, according to (\ref{scalarfactor}), 
the chiral condensate can obtain contributions only from the empty 
vertex. Using translational invariance of 
$\langle \overline{\psi} \psi (v) \rangle_{g=\infty}$ 
on the infinite lattice, one finds
(compare Eq.~\ref{7vertexz}) 
\begin{equation}
\langle \overline{\psi} \psi (v) \rangle_{g=\infty} \; = \; \frac{2}{2+m} \;
\lim_{\Lambda \rightarrow \infty} \; 
\frac{1}{Z_{g = \infty}} \; \sum_{t \in {\cal T}_7}    
\prod_{j=1}^7 (w_j)^{n_j(t)} \; \frac{n_1(t)}{|\Lambda|}.
\end{equation}
Thus in the strong coupling limit the chiral condensate is proportional 
to the density of vertex 1 of Fig.~\ref{7vertex}, i.e.~the empty tile. 
Similarly it is possible
to obtain simple expressions for other fermionic observables in the strong
coupling limit.

\section{Discussion}
In this article we have derived the loop representation for the fermion
determinant of 2-D 
Wilson lattice fermions coupled to U(1) gauge fields, i.e.~the fermion
determinant of the lattice 
Schwinger model. The construction is based on first using standard hopping
expansion which
already expresses the determinant in terms of loops which are the natural
(gauge invariant) 
variables for the fermion determinant. Standard hopping expansion, however,
only gives a loop
expansion for the free energy, and thus to obtain the desired result, the
Boltzmann factor was expanded.
An essential ingredient for this expansion was the introduction of a
locally varying mass term. For our 
problem it was shown that the determinant is a polynomial in the local mass
terms with a degree 
of two. This considerably restricts the number of terms which can appear
when expanding the
Boltzmann factor. For the remaining terms it was shown that they can be
obtained by summing over
local pieces of loops, i.e.~over the vertices given in Fig.~\ref{tiles}.

The gauge fields were integrated out and the result was expressed in terms
of occupation numbers for
the plaquettes enclosed in the fermion loops. The whole formalism
generalizes to the case of 
non-vanishing chemical potential in a natural way. Gauge field variables
can be treated on the same 
footing as the fermion loops, i.e.~by assigning occupation numbers for the
plaquettes followed by 
integrating the link variables. Local fermionic observables were generated
by differentiating the
determinant with respect to the external fields. In the loop picture their
vacuum expectation values
simply correspond to counting the net flow of fermion loops through the
sites (for scalars) and links 
(vectors) of the lattice. 
\\

Certainly the lattice Schwinger model is a relatively simple system, but
the essential ingredients of our
construction can be carried over to more realistic models. In particular, a
hopping expansion of the 
fermion determinant for Wilson fermions 
is straightforward also for higher dimensions and
non-abelian gauge groups. 
Furthermore the necessary traces over the $\gamma$-matrices can be computed
in closed form 
also in 4 dimensions \cite{St81}. It is also possible to again apply the
trick of introducing a locally 
varying mass
term and, in the region of convergence, use it to 
rule out infinitely many terms in the expansion of the
Boltzmann factor. The number of 
allowed loops running through links and sites will depend on the dimension
of the problem and the number
of colors. We expect that also in higher dimensions it is possible to
decompose the loops into local 
elements. For non-abelian models, these elements will, however, be more
involved, since also for the 
gauge fields one now has to compute traces. The inclusion of chemical
potential can be treated in 
exactly the same way as we have done for the Schwinger model. Also local
fermionic observables 
will again correspond to counting the net flow of fermion lines through
sites and links. The treatment of
the gauge field integral, however, becomes more complicated. For abelian
fields in higher dimensions 
it is still possible to build the links on our loops by filling surfaces
\cite{ArFoGa96}, and a subsequent
duality transformation (see e.g.~\cite{Sa80}) produces simple Gaussian
weights. For non-abelian 
gauge groups no such techniques are known and probably one would have to
rely on perturbative 
calculations.  

We remark that also fermionic models such as the Thirring model or the
Gross-Neveu model are 
accessible to the techniques developed here. The four-fermi
terms are then
generated by coupling
an auxiliary field to fermionic bilinears and these auxiliary fields are
subsequently integrated with
Gaussian measures. The auxiliary fields necessary for generating the above
mentioned models 
are abelian fields and also their Gaussian integrals are straightforward to
solve (see e.g.~\cite{Ga98b}
for a study of this approach to the Gross-Neveu model in 2D). 
Thus for purely fermionic
models the tools for finding
a loop representation are readily available. It is hoped that such a
representation opens new ways
for numerical simulations of these systems, in particular for the case of
non-vanishing chemical 
potential.

\vspace{10mm}
\noindent
{\sl Acknowledgement:} I have considerably profited from discussions with
Mark Alford and Uwe-Jens Wiese. My special thanks goes to Christian Lang 
who in addition to valuable comments also provided me with the results of 
his computer-algebraic evaluation of the fermion determinant on small lattices.

\end{document}